\documentclass[english,aps,prl,superscriptaddress,twocolumn,longbibliography]{revtex4-1}
\usepackage[T1]{fontenc}
\usepackage[latin9]{inputenc}
\usepackage{amssymb}
\usepackage{graphicx}
\usepackage{float}
\usepackage{esint}
\usepackage{babel}


\begin{document}

\title{Towards a global quantum network}

\author{Christoph Simon}
\affiliation{Institute for Quantum Science and Technology and Department of Physics and Astronomy, University of Calgary, Calgary, Alberta, Canada T2N 1N4}

\date{\today}
\begin{abstract}
The creation of a global quantum network is within reach combining satellite links and quantum memory based approaches. Applications will range from secure communication and fundamental physics experiments to a future quantum internet.
\end{abstract}

\maketitle

A truly global quantum network would allow the distribution of quantum states and of quantum entanglement between any two points on the earth's surface. Such a network would have many applications. The most well-known one may be quantum key distribution, which promises communication whose security is ensured by the laws of physics. There are also other quantum cryptographic paradigms such as blind quantum computing \cite{broadbent}, which allows a user to perform calculations on a quantum computer without the computer finding out what calculation is being performed, and private database queries \cite{jakobi}. Global entanglement would also allow more accurate global timekeeping \cite{komar}, with applications to navigation and earth sensing, as well as more precise telescopes \cite{gottesman} and new fundamental tests of quantum non-locality and quantum gravity \cite{rideout}. In the long term, the most important application may be the creation of a global network of quantum computers (a ``quantum internet''), which might allow computations that are beyond the reach of even the most powerful individual future quantum computers.

The long-distance transmission of quantum information will almost certainly be done via optical photons in the visible to near-infrared range, because they suffer relatively low absorption and decoherence and can be detected fairly easily. However, depending on the distance, the global quantum network will probably rely on different approaches. For distances of a few hundred kilometers (maybe up to five hundred), direct optical fiber links are likely to be the best choice. For distances in the five hundred to two thousand kilometer range, fiber-based quantum repeaters (which I will briefly describe below) may turn out to be the best option. Low-earth orbit satellite links may be the best approach for distances of a few thousand kilometers. Finally, the very longest distances beyond 5000 km or so will likely require either more distant (e.g. geostationary) satellites, quantum repeaters with low-earth orbit satellite links, or low-earth orbit satellites equipped with quantum memories, or some combination of these approaches. Quantum memories, which allow one to store and retrieve quantum information, e.g. in the form of photons in well-defined quantum states, will also be important in order to connect the ground stations of satellite links, which will likely have to be in relatively remote locations, to urban end users with minimal loss. In the following I will comment on the state of the art and future challenges for these various approaches.

Regarding ground-based links, for example, quantum key distribution has been realized over 300 km of fiber \cite{korzh}, and entanglement has been distributed over a 144 km free-space link  using a major telescope \cite{ursin}. The best currently available fibers have a loss of about 0.17 dB/km, which implies a transmission probability of $10^{-17}$ for 1000 km. This explains why classical telecommunication uses repeaters (i.e. amplifiers), and why directly sending single photons over thousands of kilometers of fiber is not an option. Unfortunately similar amplification is not possible for quantum signals because of the no-cloning theorem, so more elaborate techniques such as quantum repeaters are needed, see below.

In principle the problem could also be solved by developing fibers with significantly lower loss. However, after impressive improvements in the second half of the 20th century, silica fiber technology seems to have reached a plateau, with only modest improvements over the last two decades (from 0.2dB/km to 0.17dB/km). Regarding alternative technologies, for example, heavy-metal fluoride fibers have a lower theoretical limit, but are still two orders of magnitude more lossy than silica fibers after decades of work \cite{saad}. Free-space links using airplanes or balloons are limited by the curvature of the earth and absorption in the atmosphere.

Despite their relative complexity, quantum repeaters \cite{sangouard} are therefore likely to be an important component of the global quantum network for the foreseeable future. The basic principle of quantum repeaters is to independently create entanglement for individual links of manageable distance and store this entanglement in quantum memories. Then these links can be connected through entanglement swapping (i.e. quantum teleportation of entanglement) to create entanglement over longer distances. The quantum memories are essential because they ensure that entanglement creation doesn't need to be successful simultaneously in all individual links, which would be forbiddingly difficult because, due to the loss in the fiber and to other factors, each link only has a certain success probability for each try. This can be compared to rolling dice: it is unlikely to roll four sixes, say, at the same time. The quantum memory is equivalent to getting to ``store'' each six, while continuing to roll the remaining dice, until each one has yielded a six.

I focus here on ``first-generation'' quantum repeaters \cite{sangouard} in the terminology of Ref. \cite{muralidharan}, which do not rely on quantum error correction. Error-correction based approaches promise higher rates in the long run, but will require significantly more resources \cite{muralidharan}. Similarly, I focus on qubit-based approaches, which are currently more developed than continuous-variable schemes.

Regarding physical platforms, one can distinguish quantum memories that are realized with single quantum systems and those based on ensembles. Important single-system approaches include single atoms in high-finesse cavities \cite{ritter}, trapped ions, where the goal of connecting quantum processors provides an important motivation \cite{monroe}, and nitrogen-vacancy centers in diamond \cite{hensen}. Ensemble-based approaches include atomic vapors or cold gases \cite{yang}, as well as solid-state approaches, in particular those based on rare-earth ion doped crystals \cite{hedges,zhong,bonarota,sinclair}, which are attractive because of their potential for using many distinct frequency channels thanks to the large ratio of inhomogeneous to homogeneous broadening in these materials. 

The exact requirements for quantum memories to be useful for quantum repeaters are protocol-dependent, and there can be significant trade-offs between different metrics, e.g. storage time and multi-mode capacity. But efficiencies around 90\%, storage times in the millisecond range (where a fundamental timescale is set by the communication time for a single link, which is limited by the speed of light), and significant multi-mode capacity are all likely to be necessary for practically useful repeaters. These tentative benchmarks are within reach, but have yet to be demonstrated simultaneously in a single system.

Experimentally, memories based on cold atomic gases may currently be the most mature, having demonstrated, for example, 50\% efficiency at 50 ms storage time and efficiencies up to 76 \% for shorter times \cite{yang}. Rare-earth doped crystal based memories have demonstrated 69\% efficiency \cite{hedges} and six hour spin coherence time \cite{zhong} (however, the latter experiment did not include light storage), as well as the storage of over one thousand temporal modes \cite{bonarota} and tens of spectral modes \cite{sinclair}.

Progressing from quantum memory demonstrations to more complex quantum repeater type experiments, teleportation between atomic gas memories separated by 150 m of fiber has been demonstrated \cite{bao}, as well as teleportation of a photon onto a solid-state memory over 25 km of fiber in a laboratory environment \cite{bussieres}. A recent loophole-free test of quantum non-locality using nitrogen-vacancy centers \cite{hensen} can also be seen as the demonstration of a single quantum repeater link over a distance of 1.3 kilometers. Also important in this context, even if they don't involve quantum memories, are two recent experiments that demonstrated the feasibility of quantum teleportation with independent entanglement sources over urban fiber networks \cite{sun,valivarthi}.

The next important challenges for ground-based quantum repeaters include the demonstration of an individual repeater link over a significant distance (e.g. 100 km), the connection of two or more repeater links via entanglement swapping, and the demonstration of a quantum repeater architecture that achieves a higher entanglement distribution rate than direct transmission. A critical technology for many of these steps will be improved photonics engineering, e.g. integration to minimize coupling losses, which are often a major limiting factor in current experiments. A different challenge with great potential practical benefits is the development of quantum repeaters that can operate at room temperature. For example, nitrogen-vacancy related nuclear spins exhibit a coherence time of 1 second even at room temperature \cite{maurer}.

Despite these many impressive results, distances beyond 2000 km or so are likely out of reach for purely ground-based quantum networks for the medium term, since they would require many complex quantum repeater nodes. The development of quantum communication satellites offers an attractive solution. China has taken the lead with the recent launch of the Micius satellite, which has allowed the demonstration of direct entanglement distribution over 1200 km \cite{micius-entanglement}, as well as quantum teleportation to the satellite \cite{micius-tele} and quantum key distribution over the satellite link \cite{micius-qkd}.

Satellite links do have significant limitations, including intermittency due to orbits and weather and relatively low rates (although the latter could potentially be boosted by frequency multiplexing). As a consequence, every photon transmitted from a satellite is likely to be very valuable. Moreover the ground stations for satellite links will likely have to be located at least tens of kilometers from major population centers to avoid light and air pollution and turbulence, resulting in further significant photon loss in the final terrestrial links from the ground station to the end user.

Fortunately, the quantum repeater principle provides a potential solution for this problem. One can create and store entanglement for the final terrestrial links between ground station 1 and city 1 and between ground station 2 and city 2 respectively. When the satellite succeeds in creating entanglement between ground stations 1 and 2, entanglement between city 1 and city 2 can be created via entanglement swapping at the two ground stations. This creates effectively loss-free links between the ground stations and the cities and ensures that none of the precious photons from the satellite are wasted.

In order for this approach to work, one has to know when entanglement between the ground stations has been established. This can be achieved either by non-destructive photon detection \cite{QND}, if the source of entanglement is on the satellite and the photons are sent down following a ``downlink'' approach as in Ref. \cite{micius-entanglement}, or by teleportation of entanglement via the satellite, extending the ``uplink'' approach of Ref. \cite{micius-tele}. In either scenario the requirements on the quantum memories in the terrestrial links are more modest than for long-distance ground-based quantum repeaters, because the ``heavy lifting'' in terms of distance is done by the satellite link.

The creation of loss-free links between satellite ground stations and cities is one example for how satellites and quantum memories can complement each other, but there are many others. Applying the quantum repeater principle to multiple satellite links may be the most practical approach for creating entanglement over truly global distances \cite{boone}, see Figure 1. This is well aligned with the current trend in the space industry towards relatively cheap low-earth orbit satellites. Quantum microsatellites are already being developed \cite{takenaka}. Very long-term memories \cite{zhong} could mitigate the intermittency problem of satellite links because they would allow one to stock up on entanglement while the satellite link is available. A satellite carrying an entangled photon pair source and a long-term quantum memory would also offer an alternative approach to global quantum communication, because it could send one photon down over one location, store the other one in the memory and send it down to a different location later on.

\begin{figure}
\scalebox{0.46}{\includegraphics*[viewport=40 395 580 600]{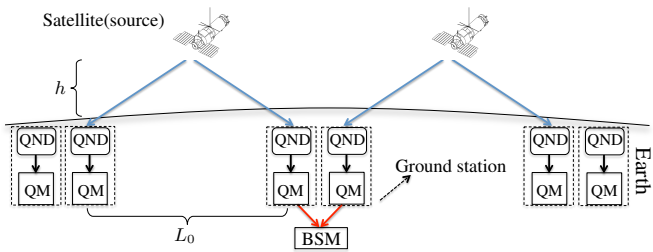}}
\caption{Quantum repeater architecture with satellite links from Ref. \cite{boone}. Each individual link(of length $L_0$) consists of an entangled photon pair source on a low-earth orbit satellite (at height $h$) and two ground stations consisting of quantum non-demolition (QND) measurement devices and quantum memories (QM). The arrival of a photon at each ground station is heralded by the QND devices, which detect the presence of a photon non-destructively and without revealing its quantum state. The entanglement is then stored in the memories until information about successful entanglement creation in a neighboring link is received. Then the entanglement can be extended by entanglement swapping based on a Bell state measurement (BSM). A small number of such links are sufficient for spanning global distances.}\label{architecture}
\end{figure}

Based on the above discussion, future important milestones involving satellite quantum links include connecting a satellite link and a ground link, connecting two or more satellite links by entanglement swapping, and generally experiments that combine satellite links and quantum memories, either on the ground or on the satellite. Realizing the vision of a quantum internet will also require interfacing promising quantum computing platforms with photons, e.g. by developing microwave-to-optical transducers for superconducting quantum processors \cite{lehnert}. There are also important quantum computer science questions, including how best to harness the power of distributed quantum computing.

The development of quantum network technology is currently seen as a priority area in most leading economies, notably in China, where, in addition to the mentioned quantum satellite, there is also work on a chain of individual quantum key distribution links stretching over 2000 km from Beijing to Shanghai. Quantum communication is a key area in the EU flagship initiative on quantum technology. I am also aware of major initiatives in the US, the UK, the Netherlands, Japan, Singapore, as well as in Canada, which has recently committed to building its own quantum communication satellite. Industrial players are also beginning to join the fray. In my personal opinion we have reached the point where the question is no longer whether we are going to have a global quantum network, but only when and how exactly we will get there. But there is still a lot of interesting work to be done, and a large part of this work will be in the domain of photonics.


\begin{thebibliography}{30}


\bibitem{broadbent} Barz, S. {\it et al.} {\it Science} {\bf 335}, 303-308 (2012).
\bibitem{jakobi}  Jakobi, M. {\it et al.} {\it Phys. Rev. A} {\bf 83}, 022301 (2011).
\bibitem{komar}  Komar, P. {\it et al.} {\it Nature Phys.} {\bf 10}, 582 (2014).
\bibitem{gottesman}  Gottesman, D., Jennewein, T. \& Croke, S. {\it Phys. Rev. Lett.} {\bf 109}, 070503 (2012).
\bibitem{rideout}  Rideout, D. {\it et al. Class. Quant. Grav.} {\bf 29}, 224011 (2012).
%\bibitem{kimble} Kimble, H.J. {\it Nature} {\bf 453}, 1023-1030 (2008).
\bibitem{korzh}  Korzh, B. {\it et al. Nature Photon.} {\bf 9}, 163 (2015).
\bibitem{ursin}  Ursin, R. {\it et al. Nature Phys.} {\bf 3}, 481 (2007).
%\bibitem{nocloning} Ortigoso, J. {\it arXiv}:1707.06910 (2017).
\bibitem{saad} Saad, M. doi:10.1117/12.915295 (2011).
%\bibitem{briegel}  Briegel, H.-J., D\"{u}r, W., Cirac, J.I., \& Zoller, P. {\it Phys. Rev. Lett.} {\bf 81}, 5932 (1998).
\bibitem{sangouard}  Sangouard, N., Simon, C., De Riedmatten, H. \& Gisin, N. {\it Rev. Mod. Phys.} {\bf 83}, 33 (2011).
\bibitem{muralidharan} Muralidharan, S. {\it et al. Sci. Rep.} {\bf 6}, 20463 (2016).
%\bibitem{dias} Dias, J. \& Ralph, T.C. {\it Phys. Rev. A} {\bf 95}, 022312 (2017).
\bibitem{ritter} Ritter, S. {\it et al. Nature} {\bf 484}, 195 (2012).
\bibitem{monroe}  Monroe, C. {\it et al. Phys. Rev. A} {\bf 89}, 022317 (2014).
\bibitem{hensen}  Hensen, B. {\it et al. Nature} {\bf 526}, 682 (2015).
%\bibitem{duan}  Duan, L.M., Lukin, M.D., Cirac, J.I., \& Zoller, P. {\it Nature} {\bf 414}, 413 (2001).
%\bibitem{afzelius} Afzelius, M., Simon, C., De Riedmatten, H. \& Gisin, N. {\it Phys. Rev. A} {\bf 79} 052329 (2009).
\bibitem{yang}        Yang, S.-J., Wang, X.-J., Bao, X.-H., \& Pan, J.-W. {\it Nature Photon.} {\bf 10}, 381-384 (2016).
\bibitem{hedges}  Hedges, M.P., Longdell, J.J., Li, Y., \& Sellars, M.J. {\it Nature} {\bf 465}, 1052-1056 (2010).
\bibitem{zhong}  Zhong, M. {\it et al. Nature} {\bf 517}, 177-180 (2015).
%\bibitem{clausen}  Clausen et al, Nature 469, 508 (2011)
%\bibitem{saglamyurek}  Saglamyurek et al, Nature 469, 512 (2011),
\bibitem{bonarota}  Bonarota, M., Le Gou\"{e}t, J.-L., Chaneli\`{e}re, T. {\it New J. Phys.} {\bf 13}, 013013 (2011).
\bibitem{sinclair}  Sinclair, N. {\it et al. Phys. Rev. Lett.} {\bf 113}, 053603 (2014).
\bibitem{bao}  Bao, X.-H. {\it et al. Proc. Nat. Acad. Sci.} {\bf 109}, 20347 (2012).
\bibitem{bussieres}  Bussi\`{e}res, F. {\it et al. Nature Photon.} {\bf 8}, 775 (2014).
\bibitem{sun} Sun, Q.-C. {\it et al. Nature Photon.} {\bf 10}, 671-675 (2016).
\bibitem{valivarthi}  Valivarthi, R. {\it et al. Nature Photon.} {\bf 10}, 676-680 (2016).
\bibitem{maurer}  Maurer, P.C. {\it et al. Science} {\bf 336}, 1283-1286 (2012).
\bibitem{micius-entanglement}  Yin, J. {\it et al. Science} {\bf 356}, 1140-1144 (2017).
\bibitem{micius-tele} Ren, J.-G. {\it et al. Nature} doi:10.1038/nature23675 (2017).
\bibitem{micius-qkd} Liao, S.-K. {\it et al. Nature} doi:10.1038/nature23655 (2017).
\bibitem{QND} Reiserer, A., Ritter, S., \& Rempe, G. {\it Science} {\bf 342}, 1349-1351 (2013).
\bibitem{boone}  Boone, K. {\it et al. Phys. Rev. A} {\bf 91}, 052325 (2015).
\bibitem{takenaka} Takenaka, H. {\it et al. Nature Photon.} {\bf 11}, 502-508 (2017).
\bibitem{lehnert} Andrews, R.W. {\it et al. Nature Phys.} {\bf 10}, 321-326 (2014).
%\bibitem{brierley}  Brierley, S. {\it arXiv}:1507.04263
\end{thebibliography}
\end{document}